# Spontaneous Breaking of CP Symmetry by Orbifold Moduli.

B. Acharya[c], D. Bailin[a], A. Love[b], W.A. Sabra[b] and S. Thomas[c]

[a] *School of Mathematical and Physical Sciences,*
*University of Sussex, Brighton BN1 9QH, U.K.*

[b] *Department of Physics,*
*Royal Holloway and Bedford New College,*
*University of London,*
*Egham, Surrey, U.K.*

[c] *Department of Physics,*
*Queen Mary and Westfield College,*
*University of London,*
*Mile End Road, London, U.K.*

## ABSTRACT

CP-violating phases which contribute to the electric dipole moment (EDM) of the neutron are considered in the context of orbifold compactification of the heterotic string. In particular, we study the situation where CP is spontaneously broken by moduli fields acquiring, in general, complex expectation values at the minimum of duality invariant low energy effective potentials. We show, by explicit minimization of such a potential in the case of the $\mathbf{Z}_6-\mathrm{IIb}$ orbifold, that it is the presence of so called Green-Schwarz anomaly coefficients $\delta^i_{\mathrm{GS}}$, that leads to significant CP violating expectation values of the moduli. By evaluating the soft supersymmetry breaking moduli dependent $A$ and $B$ terms in this model, we find that the experimental bounds $\Phi(A)$, $\Phi(B) \leq 5\times 10^{-3}$ are exceeded for a particular range of values of the auxiliary field of the $S$ modulus.

Recently there has been much interest in the problem of how the various moduli in orbifold compactifications of the heterotic string acquire non-vanishing expectation values in the low energy effective action. Promising mechanisms include hidden sector duality invariant gaugino condensation [1, 2] as well as other scenarios based on the principle of $S$-duality [3]. When the various compactification moduli acquire expectation values, they lead naturally to the appearance of various soft supersymmetry breaking operators in the low energy effective action [4]. Although in the simplified case of an overall $T$ modulus, the value at the minimum was found to be real [1], it is generally believed that complex expectation values can arise in more realistic cases. Indeed, such values were found in the context of the $\mathbf{Z}_6 - \text{IIb}$ orbifold studied in [6]. Complex values for the moduli generically lead to complex soft terms which in turn may generate observable CP violation, since it was shown in [7] that such terms contribute to the EDM of the neutron.

In [8] a model independent analysis of soft terms, including constraints based on CP violation through complex moduli values was presented. Of necessity, such an analysis parameterizes all the dynamics of the moduli, and studies the general properties of the soft terms. It was concluded that CP violation from the $A$ terms could be suppressed for large enough $f_S$ ( $f_S$ being proportional to the auxiliary field of the $S$ modulus), but that this was not necessarily the case for the phases due to $B$ terms. In this paper we want to improve on this situation by explicitly minimizing effective potentials in the $T$ and $U$ moduli and hence calculate the CP violation in a specific model. Since we do not, as yet, have a firm understanding of $S$-dynamics, we shall not commit ourselves to any particular scheme that fixes $S$. Rather like [8] we shall parameterize its dynamics. So the approach we shall adopt is to minimize the effective potential in the $T$ and $U$ moduli for a range of values of $f_S$, with the latter regarded as a free parameter. The advantages of such an approach are at least twofold. Firstly we will establish that there are indeed minima at which $T$ and $U$ moduli acquire complex values. Secondly, we can establish whether or not significant CP violation occurs from the various $B$ terms present, something that the analysis in [8] was not able to do.

To begin we remind the reader of some familiar formulae concerning the low energy $N = 1$ supergravity derived from string theory compactified on orbifolds (see for example the first reference in [4]). The Kähler potential $K$ relevant to our discussion takes the form

$$\begin{aligned} K \;=& -\ln y - \sum_i \ln (T_i + \bar{T}_i) + \sum_\alpha \prod_i (T_i + \bar{T}_i)^{-n_{\alpha i}} \phi_\alpha^* \phi_\alpha \\ &+ (\sum_{\alpha\beta} Z_{\alpha\beta}(T, \bar{T}) \phi_\alpha \phi_\beta + \text{c.c.}) + O(\phi^4) \end{aligned} \tag{1}$$

where

$$y \;=\; S + \bar{S} + \sum_i \frac{\delta^i_{\rm GS}}{4\pi^2} \ln (T_i + \bar{T}_i). \tag{2}$$

In (1) and (2), $S$ is the complex field whose real part is related to the dilaton. the sum over $i$ includes not only the three diagonal $T$ moduli, but also implicitly the $U$ moduli (the number of which is given by the Hodge number $h_{(2,1)}$). The matter fields $\phi_\alpha$ in (1) have modular weights $n_{\alpha i}$ with respect to each $SL(2, Z)$ duality group (or subgroup) acting on the $T$ and $U$ moduli. We have also allowed for off-diagonal mixing terms between certain matter fields in the Kähler potential (1) via the moduli dependent function $Z_{\alpha\beta}$. Such terms are known to lead to effective $\mu$-terms for the two Higgs multiplets of the minimal supersymmetric standard model [10], and hence to soft supersymmetry breaking mass terms. We shall return to the specific form of $Z_{\alpha\beta}$ later on, when evaluating the CP violating parts of the soft terms. The Green-Schwarz coefficients $\delta^i_{\rm GS}$ are required [11] for the cancellation of the anomalies of the underlying $\sigma$-model. In this case, for the modular transformation of the Kähler potential to be as at tree level, a non-trivial modular transformation of the dilaton field $S$ should be allowed.

In general, the effective potential $V$ is given by [1, 2]

$$\begin{aligned} y \prod_i (T_i + \bar{T}_i) V =& |W|^2 \Big\{ |f_S|^2 - 3 \\ &+ \sum_i \frac{y}{(y + \frac{\delta^i_{\rm GS}}{4\pi^2})} |1 + \frac{\delta^i_{\rm GS}}{4\pi^2 y} - \frac{\delta^i_{\rm GS}}{4\pi^2 y} f_S - (T_i + \bar{T}_i) \frac{W_{T_i}}{W}|^2 \Big\} \end{aligned} \tag{3}$$

where $W$ is the superpotential and the subscripts $S$ and $T_i$ indicate differentiation with respect to that moduli and $f_S = (1 - y\frac{W_S}{W})$ is proportional to the auxiliary field of $S$ .

Fixing the values of the moduli requires the knowledge of the non-perturbative modular invariant gaugino condensate superpotential. Given the fact that such a superpotential is generally a modular form with respect to subgroups of $SL(2,Z)$, we may express it as

$$W^{np} = e^{-\frac{24\pi^2}{b}S} \prod_{i,m} \left[\eta(\frac{T_i}{l_{im}})\right]^{-C_{im}\,(1-\frac{6\delta^i_{\rm GS}}{b})} \tag{4}$$

The coefficients $C_{im}$ and $l_{im}$ were given in [6] for various Coxeter orbifolds and $\sum_m C_{im} = 2$ for all $i$. For the pure gauge case, where there is no matter charged under the hidden sector gauge group, $b = 3C(G)$, where $C(G)$ is the Casimir for the gauge group $G$. Although (4) correctly describes the nonperturbative superpotential arising from conventional gaugino condensation [1, 2], it is clear that it does not transform as a modular form under $S$ duality. Recently superpotentials arising from $S$-duality invariant gaugino condensation have been discussed in [3], where the simple exponential $S$ dependence in (4) is replaced by modular forms of $S$. The significance of working with 1-condensate models is that the superpotential in (4) factorizes with respect to the $S$, $T$ and $U$ moduli.

Of course, by incorporating $S$ duality, the form of the $S$ dependence in (4) is not uniquely specified. For our purposes, we want to investigate possible CP violating expectation values of the $T$ and $U$ moduli, without necessarily specifying the dynamics of the $S$ field. This philosophy was emphasized in [8] where a general model-independent analysis of CP violation from such moduli was performed. What is missing from such an approach, and what we shall investigate in this paper, is how such CP violating values for the moduli actually arise, and what is their magnitude. Therefore we shall minimize the effective potential (3) in the $T$ and $U$ moduli for various fixed values of the auxiliary field $f_S$ and for Re$(S) \approx 2$. Here Re$(S) = 1/g^2_{string}$, so fixing Re$(S)$ in this way is not only consistent with a perturbative string coupling constant, but also gives $\alpha = 4\pi/g^2_{string} \approx 25$ consistent with gauge coupling constant unification. By exploring how the values of $T$ and $U$ and the resulting CP violating phases change as we vary $f_S$, we will be modelling the effects of different scenarios for $S$ dynamics. Hence we shall use a generalized 1-condensate superpotential analogous to (4) of the form

$$W^{np} = \Omega(S) \prod_{i,m} \left[ \eta(\frac{T_i}{l_{im}}) \right]^{-C_{im}\,(1-\frac{6\delta^i_{\rm GS}}{b})} \tag{5}$$

where in (5) it will not be necessary to specify the explicit form of $\Omega$, since the $\Omega$ dependence of both the potential (3) and the CP violating phases (to be discussed later) always occurs through the auxiliary field $f_S$. For an $E_8$ hidden sector condensing group, $b = 90$ in (5), whilst the Green-Schwarz coefficients $\delta^i_{\rm GS}$ have been computed in [11] for various orbifolds. Upon substitution of the $W_{np}$ of eqn. (5) into eqn (3), one finds that the function, $\hat{G}^j$ appears where the latter is related to the generalized weight two Eisenstein function $G^i_2$ which transforms in general under subgroups of $SL(2, Z)$.

$$\hat{G}^j = \frac{1}{T_j + \bar{T}_j} + \sum_m C_{jm} \frac{\partial_{T_j} \eta(\frac{T_j}{l_{jm}})}{\eta(\frac{T_j}{l_{jm}})} = \frac{1}{T_j + \bar{T}_j} + G^j_2 \tag{6}$$

$\eta(T_j)$ being the Dedekind eta-function.

Whilst the task of minimizing (3) in $T$ and $U$ moduli for various values of $f_S$, ($y$ being held fixed at $\approx 4$) has to be carried out numerically, (and in a model by model analysis), it is nevertheless instructive to study the extremization equations for these moduli analytically. We shall do this for the (albeit non-realistic) case when $\delta^i_{\rm GS} = 0$, but as we shall see this does give us some indication as to what happens in more realistic examples. The extremization equation resulting from (3) may be cast in the form

$$\begin{aligned} ({\rm Re}\tilde{G}^i)\Big[2 - |f_S|^2 - \sum_j |\tilde{G}^j|^2 + {\rm Re}\tilde{G}^i + (T_i + \bar{T}_i){\rm Re}[\tilde{G}^i_{,i}]\Big] \\ + ({\rm Im}\tilde{G}^i)\Big[{\rm Im}\tilde{G}^i + (T_i + \bar{T}_i){\rm Im}[\tilde{G}^i_{,i}]\Big] \;=\; 0 \end{aligned} \tag{7}$$

and

$$\begin{aligned} ({\rm Im}\tilde{G}^i)\Big[2 - |f_S|^2 - \sum_j |\tilde{G}^j|^2 - (T_i + \bar{T}_i){\rm Re}[\tilde{G}^i_{,i}]\Big] \\ + ({\rm Re}\tilde{G}^i)\Big[(T_i + \bar{T}_i){\rm Im}[\tilde{G}^i_{,i}]\Big] \;=\; 0 \end{aligned} \tag{8}$$

where in (7) and (8) $\tilde{G}^i = (T_i + \bar{T}_i)\hat{G}^i$, and $\tilde{G}^i_{,i}$ denotes their derivatives with respect to $T_i$. We emphasize again that the index $i$ runs over all $T$ and $U$ moduli.

The solutions to equations (7) and (8) have rather striking properties as we take $|f_S|$ to asymptotically small or large values. First consider solving these equations for large values of $f_S$. Using the property that the functions $\tilde{G}^i$ and their derivatives are bounded for reasonable ($i.e.$ $O(1)$ ) values of the moduli*, it is clear that the solutions in this limit must satisfy

$$\mathrm{Re}\tilde{G}^i = \mathrm{Im}\tilde{G}^i = 0. \tag{9}$$

Since the zeros of $\tilde{G}^i$ are in one to one correspondence with the fixed points of the corresponding duality group, we learn that the values of $T$ and $U$ that extremize the potential, approach such points asymptotically as $f_S$ grows. To be consistent with our previous statement fixed point values of the moduli should be $O(1)$ which is indeed the case for $SL(2,Z)$ and its congruence subgroups [6].

We have established that for such values of $f_S$, the moduli take on their fixed point values, which in general are complex and hence might be expected (and is generally believed) to lead to CP violation. We shall see later on that, surprisingly, this is not the case if the moduli take on precisely their fixed point values.

Having discussed the solutions of (7) and (8) for large values of $f_S$, we can consider the other extreme, when $f_S$ is small, say of $O(1)$. First note that if the values of the moduli are real at an extremum of $V$, then clearly $\mathrm{Im}\tilde{G}^i = \mathrm{Im}\tilde{G}^i_{,i} = 0$, and equation (8) is automatically satisfied, whilst the second term in square brackets in eqn. (7) is also vanishing. Thus such real moduli could be minima if cancellations occur between the various terms in the first square bracket of (7). This is now plausible since all these terms will be of $O(1)$, if once more we consider reasonable values for the moduli. Although what we have stated is not a proof based on the nature of eqns (7) and (8), numerical minimization in specific models for small values of $f_S$ confirms that this is precisely what happens, $i.e.$, for such values of $f_S$, the moduli will be real at the minimum of $V$ and hence there will be no CP violation in this case. This property has already been discussed in the literature [1] where minimization of the effective potential for the case of an overall $T$ modulus and $\delta^i_{\mathrm{GS}} = 0$ lead to the value $T = 1.23$ for $f_S = 0$.

∗ a fact which may easily be verified by numerical plots of $\tilde{G}^i$ and $\tilde{G}^i_{,i}$

To illustrate the connection between the values of $f_S$ and the moduli when $\delta^i_{\rm GS}=0$, we shall consider the $\mathbf{Z}_6-{\rm IIb}$ orbifold. Although this model strictly speaking has some $\delta^i_{\rm GS}\neq 0$, we truncate the effective potential to the $\delta^i_{\rm GS}=0$ case. Later on we will relax this condition and consider the effects that non-vanishing $\delta^i_{\rm GS}$ have in the same model. There are two $T$ moduli $T_1$ and $T_3$ and a single $U$ modulus $U_3$ that have nontrivial dependence in the effective potential obtained from gaugino condensation. The associated target space duality groups are [6] $\Gamma^0_{T_3}(3)$, $\Gamma^0_{U_3-2i}(3)$, and $SL_{T_1}(2,Z)$ respectively, where

$$\Gamma^0(n) \qquad : \begin{pmatrix} a & b \\ c & d \end{pmatrix}, \qquad\qquad ad-bc=1, \quad b=0 \pmod{n}. \qquad (10)$$

In Fig.1a and Fig.1b (the solid dots) we have plotted the real and imaginary parts of the modulus $T_3$ against $f_S$ for values of $f_S$ between $(0,10)$, for the realistic case when $\delta^i_{\rm GS}\neq 0$. The plots of these moduli for $\delta^i_{\rm GS}=0$ are very similar in appearance so for simplicity we have not displayed them, but nevertheless we shall describe what happens. First, there is a unique (up to $\Gamma^0_{T_3}(3)$ duality transformations) fixed point of $\Gamma^0_{T_3}(3)$, namely $T_3=\sqrt{3}(1+i\sqrt{3})/2$, which is indicated by the dashed lines in Figs. 1a,1b. It is clear from Figs.1 that $T_3$ tends rather rapidly to this value as $f_S$ increases. Moreover, we see that for $f_S\leq 2.5$, $T_3$ becomes real. What is interesting, and apparent in Figs.1 is that $T_3$ does not smoothly interpolate between these extremes, and there is a jump around $f_S\approx 2.5$ from real to fixed point values. We have checked that for $f_S\geq 2.5$, they are within $10^{-6}$ of their fixed point values. We will see later that this rather peculiar behaviour has important consequences for CP violation.

Let us now return to the (more realistic) case when $\delta^i_{\rm GS}\neq 0$. Since $\delta^i_{\rm GS}$ appears in all formulae divided by at least a factor of $4\pi^2$ and since the maximum value of $|\delta^i_{\rm GS}|$ is 5 in $\mathbf{Z}_6$ orbifolds (for pure gauge hidden sector) [11], it would seem that including $\delta^i_{\rm GS}$ amounts to small corrections. Whilst this is indeed the case, it does not diminish the importance of such terms in possibly contributing to CP violating phases, since the experimental bounds on the latter (for example coming from soft supersymmetry breaking terms ) are of the order of $10^{-3}$ [7]. Although one can derive equations analogous to (7), (8) taking into account non-vanishing $\delta^i_{\rm GS}$, they are considerably more complicated. Nevertheless we can still expect a similar relationship between the moduli values at the minimum and the value of $f_S$, albeit slightly modified by the presence of $\delta^i_{\rm GS}$. The plots in Figs.1a,b (solid dots), illustrate this modification, again with reference to the $T_3$ modulus of the $\mathbf{Z}_6-$IIb orbifold. The main plots show the jump from real to fixed point values as described above. However the important point to notice (as illustrated in the sub-plots in Figs.1a,b which show a magnified view around the fixed points) is that there are values of $f_S$ for which the $T_3$ modulus is very close to, but not actually at it's fixed point value. In fact there is a kind of damped oscillation about the fixed points with maximum amplitude roughly $10^{-3}$, compared with the value $10^{-6}$ for the $\delta^i_{\rm GS}=0$ case. Similarly for small values of $f_S$, the modulus now develops a small imaginary part (typically of order $10^{-3}$). We shall discover in what follows that although such differences appear insignificant, they are responsible for significant CP violation. Before proceeding, we should make a few comments concerning the $U_3$ and $T_1$ moduli for non-vanishing $\delta^i_{\rm GS}$. The values of $U_3$ at the minimum are very similar to those of $T_3$ plotted in Figs1. $T_1$ on the other hand exhibits a different behaviour than that described so far. Rather than being driven close to fixed point values, we find that as $|f_S|$ increases, $T_1$ approaches the value $1.4+0.5i$, which is relatively far from the fixed points of $SL(2,Z)$. However, in this model $\delta^i_{\rm GS}$ is non-vanishing only in the complex plane associated with $T_1$, so it is perhaps not surprising that this modulus behaves differently in this respect.

It is well known that superstring theory (and hence the low energy effective action derived from it ) does not explicitly break CP symmetry, but this of course does not exclude spontaneous breaking. This is precisely what can happen when the various moduli of compactification (like $T$ and $U$ in our previous discussion ) acquire expectation values. CP breaking phases arise from the various soft supersymmetry breaking terms produced when supersymmetry is broken by the moduli vacuum values [8, 9]. The soft terms that are important in CP violation include gaugino masses, trilinear scalar couplings ('$A$' terms ) as well as bilinear scalar masses ('$B$' terms ). We shall consider the following superpotential present in the low energy supergravity theory,

$$W = W^{np} + \tilde{W} = W^{np} + \sum_{\alpha\beta\gamma} h_{\alpha\beta\gamma}\phi_\alpha\phi_\beta\phi_\gamma + \sum_{\alpha\beta}\mu_{\alpha\beta}\phi_\alpha\phi_\beta \tag{11}$$

where in eqn (11), $W^{np}$ is the duality invariant nonperturbative superpotential defined in (5), $h_{\alpha\beta\gamma}$ are yukawa couplings of the observable sector matter fields and $\mu_{\alpha\beta}$ is the so called $\mu$-term [10]. If we consider as observable sector matter that of the minimal supersymmetric standard model, then the only allowable $\mu$ terms are $\mu_{12}H_1H_2$ where $H_1$ and $H_2$ are the two electro-weak Higgs doublets. The general $A$ term arising from the effective potential (3) with superpotential (11) is

$$
\begin{aligned}
&\sum_{\alpha\beta\gamma} A_{\alpha\beta\gamma}\phi_\alpha\phi_\beta\phi_\gamma + \sum_{\alpha\beta\gamma} A'_{\alpha\beta\gamma}\bar{\phi}_\alpha\phi_\beta\phi_\gamma \;=\; m_{3/2}\frac{\bar{W}^{np}}{|W^{np}|}\{\tilde{W}^{(3)}\bar{f}_S + \\
&\sum_i \frac{y}{(y+\frac{\delta^i_{\mathrm{GS}}}{4\pi^2})}(1+\frac{\delta^i_{\mathrm{GS}}}{4\pi^2 y} - \frac{\delta^i_{\mathrm{GS}}}{4\pi^2 y}\bar{f}_S - (T_i+\bar{T}_i)\frac{\bar{W}^{np}_{\bar{T}_i}}{\bar{W}^{np}})\Big(\tilde{W}^{(3)} - \\
&\sum_\alpha \tilde{W}^{(3)}_{\phi_\alpha}\phi_\alpha n_{\alpha i} - (T_i+\bar{T}_i)\tilde{W}^{(3)}_{T_i}\Big) + \\
&\{\bar{H}_2\tilde{W}^{(3)}_{H_1} + \bar{H}_1\tilde{W}^{(3)}_{H_2}\}\Big((\bar{Z}+\bar{\mu}/\bar{W}^{np})(T+\bar{T})(U+\bar{U}) - \\
&(T+\bar{T})^2(U+\bar{U})\frac{\partial\bar{Z}}{\partial T}((T+\bar{T})\frac{\bar{\bar{W}}^{np}_{\bar{T}}}{\bar{\bar{W}}^{np}} - 1) - \\
&(U+\bar{U})^2(T+\bar{T})\frac{\partial\bar{Z}}{\partial U}((U+\bar{U})\frac{\bar{\bar{W}}^{np}_{\bar{U}}}{\bar{\bar{W}}^{np}} - 1)\Big)\}
\end{aligned}
\qquad (12)
$$

where in (12) $\tilde{W}^{(3)}$ is the cubic part of the matter superpotential given in (11) and $m_{3/2} = \mathrm{e}^{K/2}|W^{np}|$. Note that the first of the $A$ terms in (12) is holomorphic in the matter scalars $\phi_\alpha$, whilst the second ($A'$ ) involves an exchange between $\phi$ and $\bar{\phi}$, when $\phi$ is identified with either of the two doublets $H_1$ and $H_2$. Such $A'$ terms are a consequence of the Kähler mixing term between the Higgs doublets in (1). Also as noted in [10], the coefficient of this mixing $Z = Z_{12}$ only depends on the $T$ and $U$ modulus of the 2-dimensional plane associated with the fields $H_1$ and $H_2$. We should add that there are normally scaling factors in the soft supersymmetry breaking terms like those of (12) when expressed in terms of physical fields having canonical kinetic energies. We have not explicitly included them here as they constitute overall real factors, and do not contribute to CP violation.

The gaugino masses are given by

$$
m_a = \frac{m_{3/2}}{2} (\mathrm{Re} f_a)^{-1} \Big[ -y\, \bar{f}_S + \sum_i \frac{y}{(y + \frac{\delta^i_{\mathrm{GS}}}{4\pi^2})} \, (T_i + \bar{T}_i)^2 \, |\hat{G}^i|^2 \, (1 + \frac{\delta^i_{\mathrm{GS}}}{4\pi^2} \frac{\bar{W}^{np}_{\bar{S}}}{\bar{W}^{np}}) \, (\frac{\delta^i_{\mathrm{GS}}}{4\pi^2} - \frac{b^i_a}{4\pi^2}) \Big]. \tag{13}
$$

It is well known that one source of CP violation is through complex fermion masses. However, it is evident from eqn (13) that the only source of complex gaugino masses is through the $S$ field independent of the particular values of the $T$ and $U$ moduli. Although one might imagine that the $S$ field could acquire a complex expectation value, all known mechanisms give rise to real values only, including scenarios based on $S$-duality . In modelling the dynamics of the $S$-field, we shall only consider real values of $S$ and its corresponding auxiliary field $f_S$, and so we can rule out any possible CP violation from gaugino masses.

Next we consider the possible $B$-term soft supersymmetry breaking scalar masses. Again, in the context of the minimal supersymmetric standard model, the $B$-term relevant to CP violation via the EDM of the neutron, will only involve mixing between $H_1$ and $H_2$. Such a $B$-term is defined as

$$
\{m^2 \, H_1 H_2 \, + \, \mathrm{c.c.}\} \, = \, \{\mu \, B \, H_1 H_2 \, + \, \mathrm{c.c}\} \tag{14}
$$

where in (14) the definition of $B$ already takes into account the well known rescaling that the matter superpotential $\tilde{W}$ of (11) acquires in passing from the low energy supergravity theory derived from string compactification to the spontaneously broken globally supersymmetric theory, *i.e.*,

$$
\tilde{W} \, \rightarrow \, \hat{W} \, = \, \frac{W^{np}}{|W^{np}|} \tilde{W}. \tag{15}
$$

From the definition (14) we see that $B$-terms are associated with $\mu$-terms and vice-versa. There are a number of different mechanisms in the context of superstring compactification that can give rise to $\mu$-terms [10]. Such mechanisms differ in that $\mu$ is either explicitly generated in the superpotential (11) or it can be effectively induced by generating soft mass terms like those on the left hand side of (14). We shall consider examples of both mechanisms in this paper.

The first, and perhaps simplest method for generating $\mu$ is as a direct result of hidden sector gaugino condensation, when Kähler mixing through the $Z$ parameter is allowed (see eqn (1) ). As shown in [10], such terms lead to corrections to the non-perturbative superpotential quadratic in $H_1$ and $H_2$, with a moduli dependent coefficient that is essentially determined by target space duality invariance. To be specific, it is known that at tree level, the Kähler potential of the $T$ and $U$ moduli associated with the matter fields $H_1$ and $H_2$ takes the form [10]

$$K^{\rm tree} = -\ln\Big[(U+\bar{U})(T+\bar{T}) - (H_1+\bar{H}_2)(\bar{H}_1+H_2)\Big]. \tag{16}$$

$K^{\rm tree}$ is invariant under the target space duality transformations in which the fields $H_1$ and $H_2$ transform with weight 1 under $SL_T(2,Z)$

$$\begin{aligned} T &\to \frac{aT-ib}{cT+id} & \qquad U &\to U - \frac{ic}{icT+d}H_1H_2 \\ H_1 &\to (icT+d)^{-1}H_1 & \qquad H_2 &\to (icT+d)^{-1}H_2 \end{aligned} \tag{17}$$

A similar set of transformations exist corresponding to $SL_U(2,Z)$. By expanding the Kähler potential (16) in powers of $H_1\,H_2$ we can then obtain the tree level expression for $Z$ by comparison with (1). Together with 1-loop string contributions the complete $Z$ function is given by

$$Z(T,U) \;=\; \frac{1}{(T+\bar{T})(U+\bar{U})} + \frac{b^{N=2}}{16\pi^2}\hat{G}(T)\hat{G}(U), \tag{18}$$

where the functions $\hat{G}$ in (18) which were defined in (6) for a general complex plane, depend on the $T$ and $U$ moduli only, and transforms in general under a subgroup of the $SL(2,Z)$ duality group. In (18) $b^{N=2}$ is the $N=2$ supersymmetric beta function of that gauge group which rotates the matter fields $H_1$ and $H_2$.

In the $\mathbf{Z}_6-$IIb orbifold, it is the $Z_2$-plane moduli $T_3$ and $U_3$ that are associated with $H_1$ and $H_2$ and so the relevant duality groups are not $SL(2,Z)$ but rather $\Gamma^0_{T_3}(3)\times\Gamma^0_{U_3}(3)$. In [10] the $\mu$ term obtained from gaugino condensation was invariant under $SL_T(2,Z)\times SL_U(2,Z)$. The appropriate generalization to the

case at hand is

$$\mu_{cond}(S,T_1,T_3,U_3) \;=\; -2\Omega(S)\frac{1}{\eta(T_3)\eta(\frac{T_3}{3})\eta(U_3)\eta(\frac{U_3}{3})}\times \\ \partial_{T_3}\{\ln(\eta(T_3)\eta(\frac{T_3}{3})\}\partial_{U_3}\{\ln(\eta(U_3)\eta(\frac{U_3}{3})\}W'^{np}(T_1) \tag{19}$$

where in (19) $W'^{np}(T_1)$ is that part of the nonperturbative superpotential which depends on the moduli other than $T_3,U_3$ *i.e.*, $T_1$ in this case, and is given by

$$W'^{np}(T_1) = \frac{1}{\Big[\eta(T_1)\Big]^{2(1-6\frac{\delta^1_{\rm GS}}{b_{E_8}})}}, \tag{20}$$

with $b_{E_8}$ the beta function coefficient for a pure gauge hidden sector. The corresponding $B$-term (eqn (14) ) is given by

$$B_{cond} \;=\; m_{3/2}\Big[\sum_i(\rho_i\bar{\hat{G}}^i + \frac{\gamma_i}{(T_i+\bar{T}_i)} + \frac{\tilde{\delta}^i_{\rm GS}}{(T_i+\bar{T}_i)}(\frac{1-\bar{f}_S}{y}))(\frac{\tilde{\delta}^i_{\rm GS}}{(T_i+\bar{T}_i)}(\frac{1-f_S}{y})- \\ \frac{1}{\mu_{cond}}\frac{\partial\mu_{cond}}{\partial T_i} + \frac{1}{(T_i+\bar{T}_i)} - 2\frac{n_i}{(T_i+\bar{T}_i)})\frac{(T_i+\bar{T}_i)^2 y}{y+\tilde{\delta}^i_{\rm GS}} + |f_S|^2 - 1\Big] \tag{21}$$

where $\tilde{\delta}^i_{\rm GS} = \frac{\delta^i_{\rm GS}}{4\pi^2}$, and $\rho_i$ and $\gamma_i$ are constants that depend on the particular orbifold. One finds for $\mathbf{Z}_6-$IIb that (with $b_{E_8}=+90$), $\rho_3=1$, $\gamma_3=0,\rho_1=2/3,\gamma_1=1/3$, and $\delta^3_{\rm GS}=0,\delta^1_{\rm GS}=5$. Since $H_1$ and $H_2$ are untwisted matter fields associated with the $i=3$ plane, $n_1=n_2=0$, and $n_3=1$.

Finally, the other source of $B$-terms comes from the effective $\mu$ term associated with the Kähler mixing terms between $H_1$ and $H_2$ in (1). This can be obtained straightforwardly from the corresponding higgsino masses

$$\mu_{eff} = |W^{np}|\Big[Z + \sum_{T_i,U_i}(T_i+\bar{T}_i)^2\hat{G}^i\frac{\partial Z}{\partial\bar{T}_i}\Big], \tag{22}$$

The resulting $B$-term ( which we denote by $B_Z$) takes the form

$$B_Z = -2m_{3/2}\Big[\sum_i(T_i+\bar{T}_i)\bar{\hat{G}}(T_i) - 1 \\ -\frac{|W^{np}|}{2\mu_{eff}}\Big\{\sum_{i,j} -\frac{\partial^2 Z}{\partial\bar{T}_i\partial T_j}\hat{G}^i\bar{\hat{G}}^j(T_i+\bar{T}_i)^2(T_j+\bar{T}_j)^2 - \\ \sum_i\frac{\partial Z}{\partial\bar{T}_i}\hat{G}^i(T_i+\bar{T}_i)^2 - \sum_i\frac{\partial Z}{\partial T_i}\bar{\hat{G}}^i(T_i+\bar{T}_i)^2\Big\}\Big]. \tag{23}$$

It should be noted that in eqns (22) and (23), the sums over $T_i,U_i$ moduli only includes those of the $Z_2$ plane associated with the untwisted matter fields $H_1$ , $H_2$ *i.e.*, the moduli $T_3$ and $U_3$ in the case of $\mathbf{Z}_6 - \mathrm{IIb}$. In addition, we have dropped the 'cosmological constant' term in eqn (23) proportional to $V$ evaluated at its minimum . Whilst there may or may not be Kähler mixing present, it has been argued from the point of view of duality invariance of the effective potential [10] that $B_{cond}$ always exists in the presence of gaugino condensation. Hence the complete $B$ term in the presence of Kähler mixing, $\tilde{B}_Z$, is given by

$$\tilde{B}_Z = \frac{(\mu_{cond}B_{cond} + \mu_{eff}B_Z)}{\mu_{cond} + \mu_{eff}} \tag{24}$$

Given the various soft supersymmetry breaking parameters of eqns (12),(21) and (23), their contribution to the EDM of the neutron has been estimated in [7], within the context of the minimal supersymmetric standard model, and in the approximation where only the top Yukawa coupling is considered. As noted in [7], there are various unobservable phases in the soft terms mentioned, but an upper bound was obtained on the physical CP violating phases $\phi(\hat{A})$, $\phi(B)$ through the observed EDM of the neutron

$$\phi(\hat{A})\,,\phi(B) \,\leq\, 5\times 10^{-3}, \tag{25}$$

where $\phi(\hat{A}), \phi(B)$ are the phases of the $\hat{A}$ and $B$ terms, with $\hat{A}$ being the cubic coupling between $H_1, H_2$ to the top generation squarks. An important observation is that because of the rescaling of the matter superpotential (see eqn (15) ) the phase factor $\frac{\bar{W}^{np}}{|W^{np}|}$ in the $A$ terms is unobservable and may be rotated away (we saw the same phenomenon in the $B$ terms defined above.) $\hat{A}$ denotes the $A$ term with such phases removed. Consequently, it follows from (5) that the dependence of $\hat{A}$ on the $T$ and $U$ moduli is always either proportional to the modified Eisenstein function $\hat{G}^i$ or the real parts of the moduli. This is also true of the $\mu$ terms in eqs. (19) and (22) and the $B$ terms in eqns (21) , (23) and (24). So we arrive at an interesting conclusion; namely that the only way CP violating phases can be generated through expectation values of the $T$ and $U$ moduli is through the functions $\hat{G}^i$. We can now immediately see the significance of the moduli taking on their fixed point values at a minimum of the effective potential, since precisely at these points $\hat{G}^i = 0$ and there is strictly no CP violation.

More realistically, if the moduli take on complex values then a possible measure of the degree of CP violation will be how far from the fixed point such values lie. The issue of how much CP violation occurs for a given 'distance' from the fixed points really depends on the properties of the functions $\hat{G}^i$, and the exact dependence of the relevant $B$ terms on these quantities. One can imagine expanding $\hat{G}^i$ about their fixed point values, so that for moduli close to these points (which we have already seen is the case for both $\delta^i_{\rm GS}$ vanishing and non-vanishing, for sufficiently large $f_S$ ), we may truncate to linear order. Roughly speaking then, CP violation should at the very least be of a similar order to the distance of the moduli from their fixed point values. This observation is explicitly verified in Fig. 2a , where the phases $\phi(\hat{A})$ and $\phi(B_{cond})$ are plotted against $f_S$ for the $\mathbf{Z}_6-$IIb orbifold, with $\delta^i_{\rm GS} \neq 0$. The solid curves represent $\phi(\hat{A})$ whilst the dashed curves $\phi(B_{cond})$. It can be seen that the maximum CP violation in this case is of order $10^{-3}$ (with $f_S \approx 2$ ) just as we anticipated from the proximity of the moduli to the fixed points. (Figs 1a,1b ) at this value of $f_S$ . In Fig. 2b on the other hand, we find an order of magnitude larger value for $\phi(\tilde{B}_Z)$ which violates the bounds given in (25). This may be understood from the form of eqn (24) relating $\tilde{B}_Z$ to $B_{cond}$ and $B_Z$ . The imaginary part of $\tilde{B}_Z$ receives contributions proportional to the real part of $B_{cond}$ which increases like $|f_S|^2$ with increasing $f_S$. (The real part of $\tilde{B}_Z$ receives similar contributions.) Hence even though by our previous arguments we might expect CP violation of order $10^{-3}$, this can be amplified in the case of the phase $\phi(\tilde{B}_Z)$.

Having said that, we must not forget the $SL(2,Z)$ $T_1$ modulus which contributes to both the $A$ and $B$ terms, but whose values at the minimum, as noted earlier, can be quite far from the fixed points of $SL(2,Z)$. How can we reconcile this with the relative smallness of the CP violating phases in Fig 2a, 2b ? The answer is to be found in the fact that whilst it is sufficient to have moduli very close to their fixed points in order to generate small ${\rm Im}\hat{G}^i$, it is by no means necessary. It turns out that ${\rm Im}\hat{G}^i(T)$ has the property that it is strongly suppressed if ${\rm Re}T$ increases beyond its fixed point values. In the present example ${\rm Re}T_1 \approx 1.4$ at the minimum, for the range of $f_S$ considered, whereas the fixed point values are$\sqrt{3}/2$ and 1 respectively. A numerical investigation of ${\rm Im}\hat{G}^1(T_1)$ shows that such a relatively large deviation still gives ${\rm Im}\hat{G}^1(T_1) \leq 10^{-3}$, consistent with our results in Fig 2a . Thus there are in fact two 'mechanisms' present which contrive to suppress CP violation among the three moduli $T_1, T_3$ and $U_3$.

In conclusion we see that at least for the specific model considered, Kähler mixing terms of the type considered in (1) generate unacceptably large CP violating angles for $|f_S| > 3$. This is in contrast to the case without Kähler mixing, where CP violation is within experimental bounds. Given that the maximum value of $\phi(\hat{A})$ and $\phi(B)$ just falls within the experimental upper bound (25) for other values of $f_S$, it would certainly be worthwhile estimating the CP violating phases in other orbifolds [13] . At the same time a further investigation is needed into the analytic properties of the extremization equations (7) and (8) to understand the apparent jump in the moduli values. Finally, it would be interesting to include continuous Wilson line moduli in a discussion of CP violation, especially since major progress has recently been made in understanding the contribution of the former to the low energy effective potential [14].

**ACKNOWLEDGEMENT**

This work is supported in part by P.P.A.R.C. and the work of S. Thomas is supported by the Royal Society.

## Figure Captions

**Fig**. 1a. The dependence of Re$T_3$ on $f_S$ in the $\mathbf{Z}_6 - \mathrm{IIb}$ orbifold, when $\delta^i_{\mathrm{GS}} \neq 0$ (solid dots). The dashed line indicates the real part of the fixed point of the duality subgroup $\Gamma^0_{T_3}(3)$ of $SL(2, Z)$

**Fig**. 1b. The dependence of Im$T_3$ on $f_S$ in the $\mathbf{Z}_6 - \mathrm{IIb}$ orbifold, when $\delta^i_{\mathrm{GS}} \neq 0$ (solid dots). The dashed line indicates the imaginary part of the fixed point of the duality subgroup $\Gamma^0_{T_3}(3)$ of $SL(2, Z)$

**Fig**. 2a. The CP violating phases $\phi(\hat{A})$ (solid curve ) and $\phi(B)$ (dashed curve) are plotted against $f_S$, where $B = B_{cond}$, is due purely to the explicit $\mu$ term obtained from gaugino condensation.

**Fig**. 2b. The CP violating phases $\phi(\hat{A})$ (solid curve) and $\phi(B)$ (dashed curve) are plotted against $f_S$, where $B = \tilde{B}_Z$ is due to both gaugino condensation and Kähler mixing.

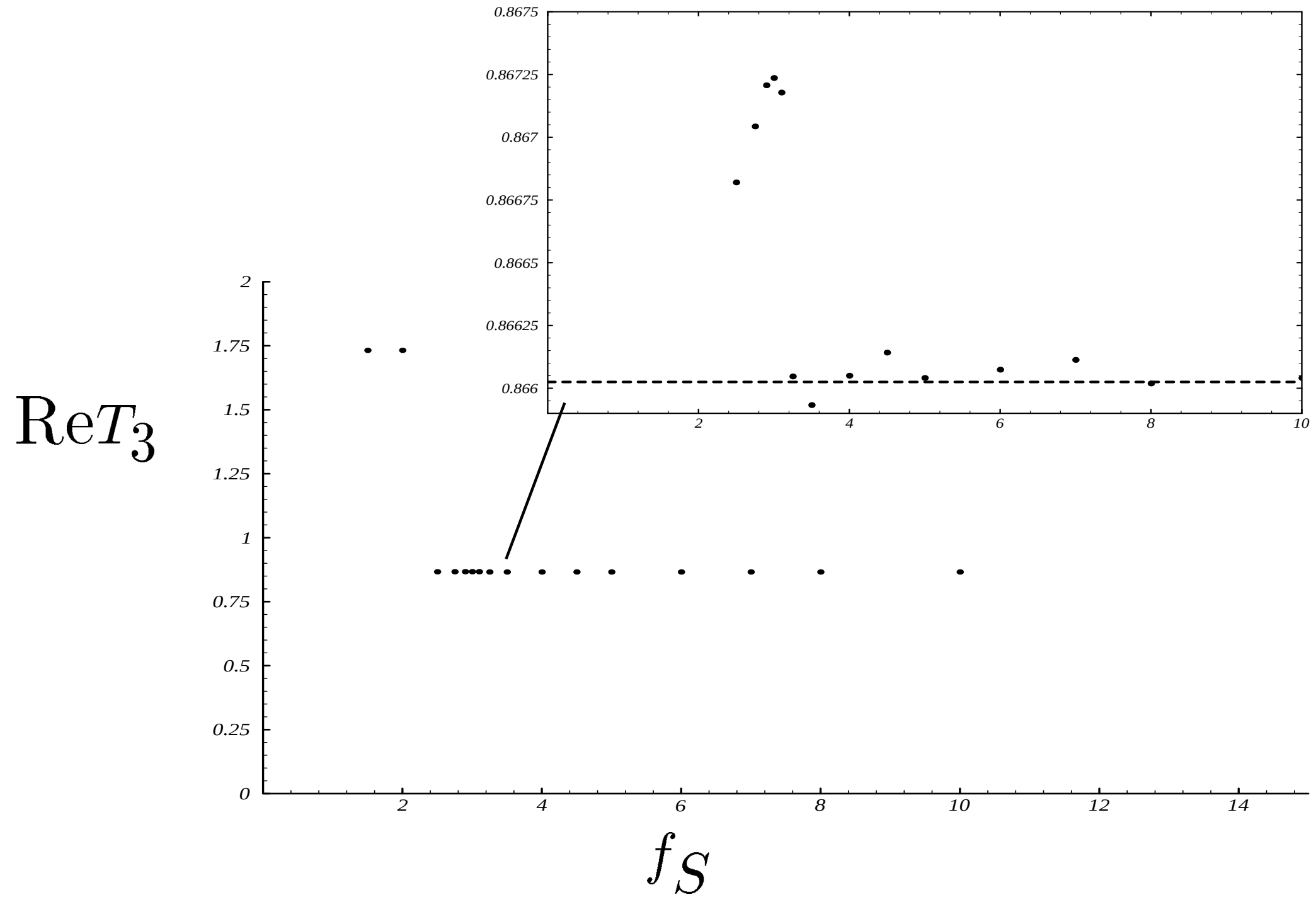

**Fig 1a**

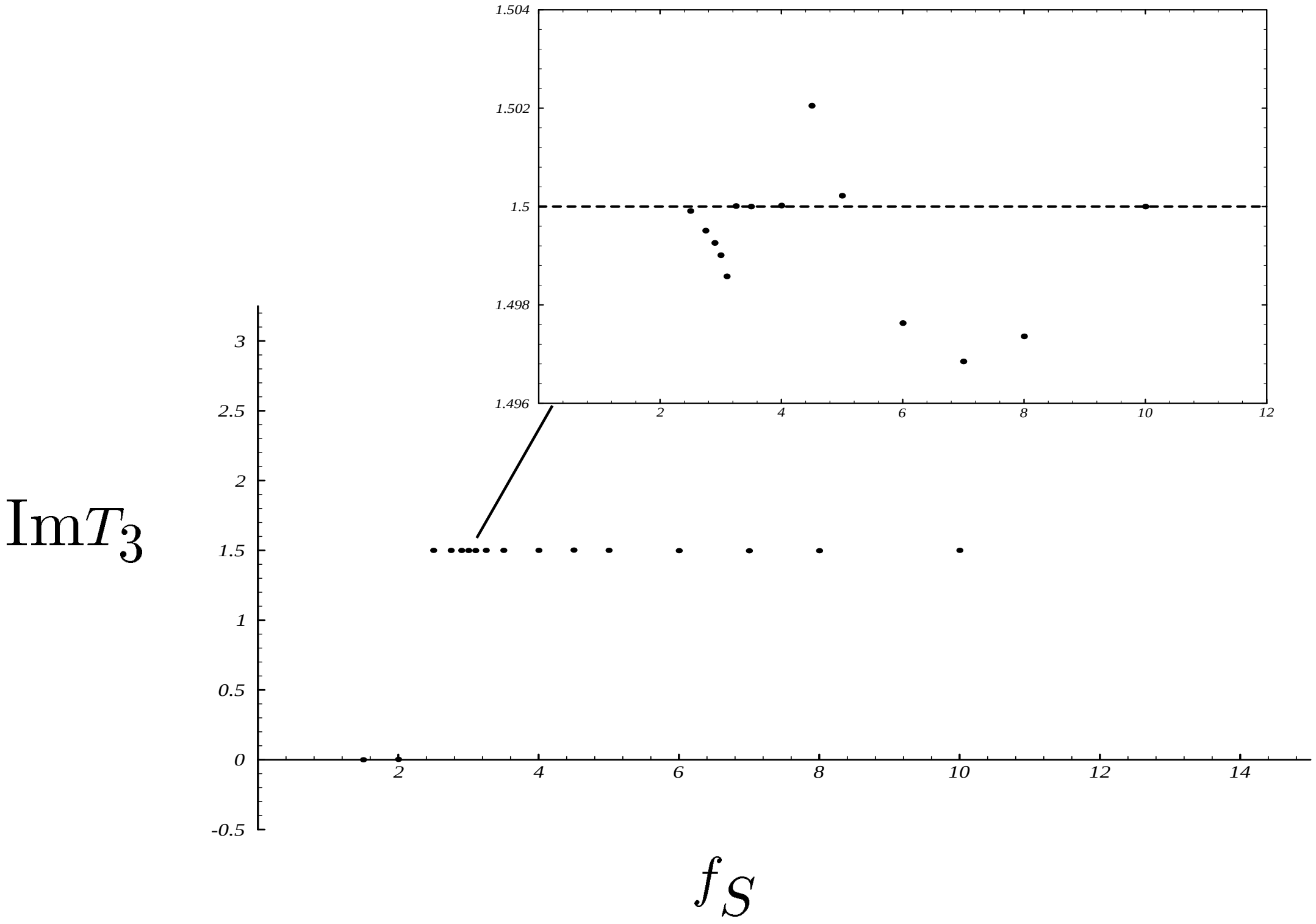

**Fig 1b**

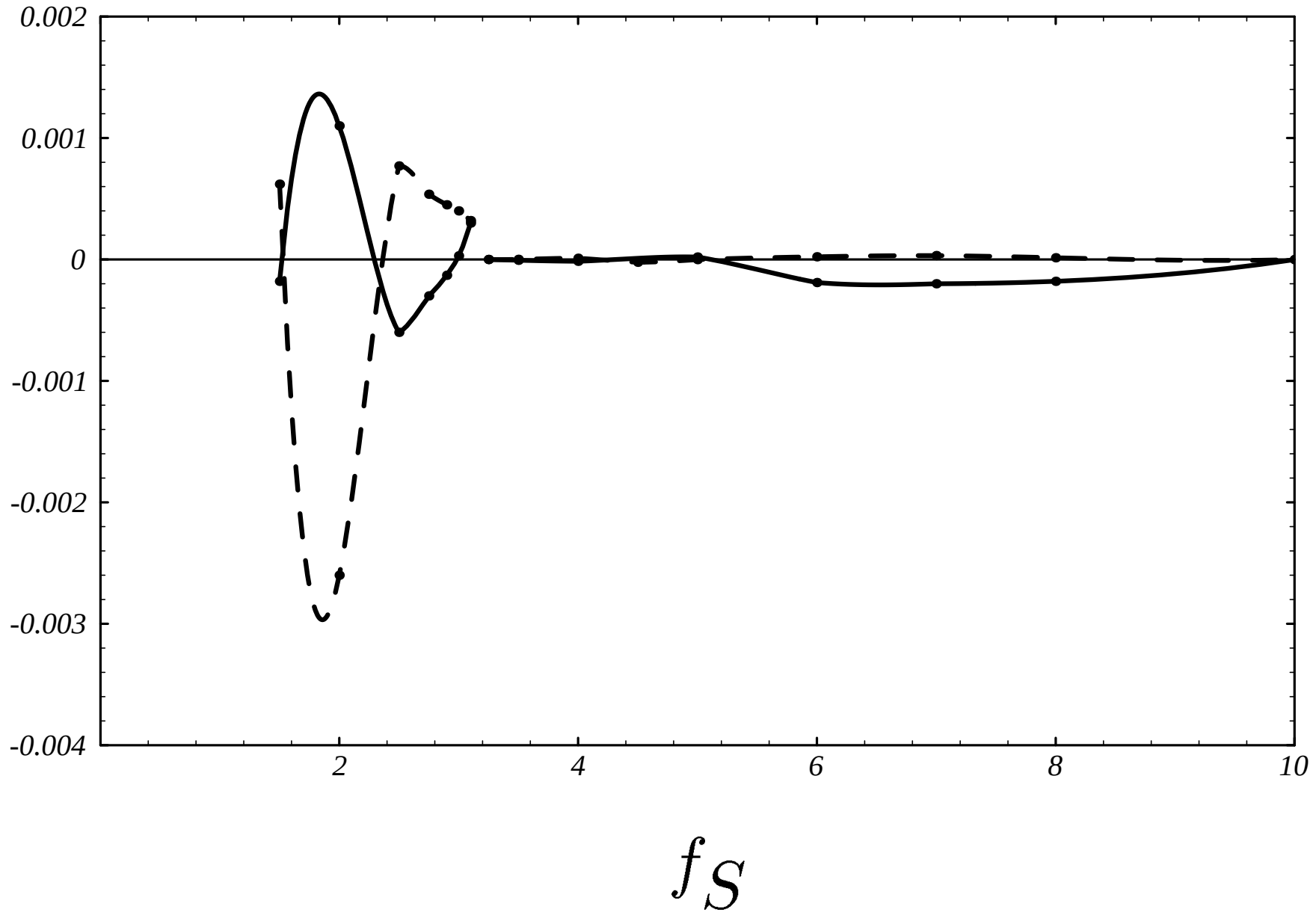

**Fig 2a**

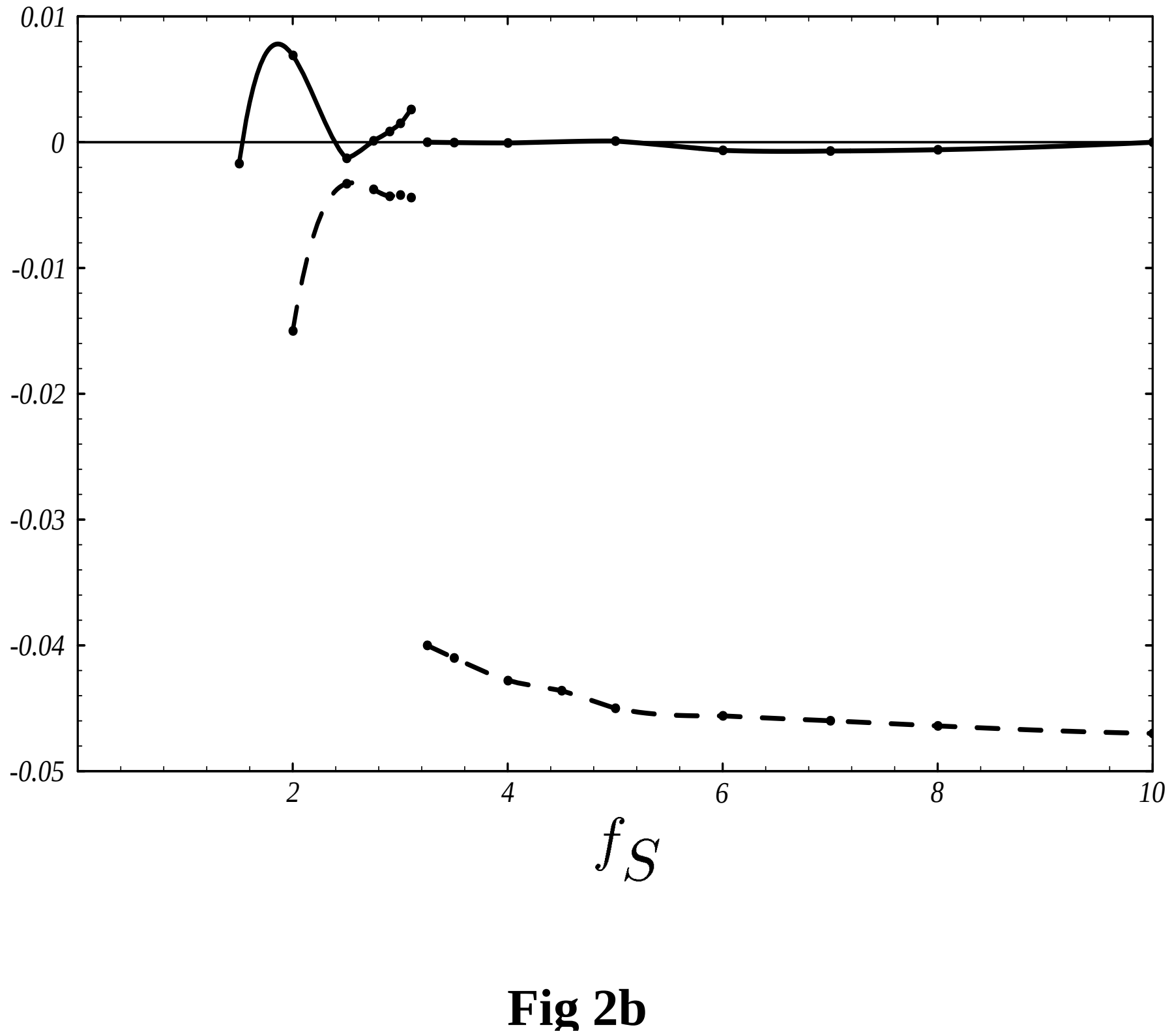

**Fig 2b**